\documentclass[letterpaper,twocolumn,10pt]{article}
\usepackage{usenix2019_v3}

\usepackage{tikz}
\usepackage{amsmath}

\usepackage{filecontents}
\usepackage{xspace}

\usepackage{breakurl}

\usepackage{tabularx}
\usepackage{booktabs}

\newcommand{\benchmark}{\textsc{DriveBench}\xspace}
\newcommand{\tool}{\textsc{AutoDriver}\xspace}

\usepackage{pgfplots}
\usepackage{wheelchart}
\usepackage{siunitx}
\definecolor{Cerulean}{RGB}{0,123,167}   
\definecolor{Green}{RGB}{0,150,0}       
\definecolor{Red}{RGB}{190,30,45}   
\definecolor{lightgray}{gray}{0.9}
\definecolor{darkgray}{gray}{0.7}
\definecolor{lightgreen}{RGB}{200,255,200}
\definecolor{greenish}{RGB}{180,255,180}
\definecolor{darkgray}{gray}{0.75}
\definecolor{lightgray}{gray}{0.9}
\definecolor{highlight}{RGB}{180,255,180}
\renewcommand{\arraystretch}{1.25}
\definecolor{catrow}{RGB}{235,243,252}
\definecolor{darkgray}{gray}{0.75}
\definecolor{lightgray}{gray}{0.9}
\definecolor{highlight}{RGB}{180,255,180}
\definecolor{codegreen}{rgb}{0,0.6,0}
\definecolor{codegray}{rgb}{0.5,0.5,0.5}
\definecolor{codepurple}{rgb}{0.58,0,0.82}
\definecolor{backcolour}{rgb}{0.95,0.95,0.92}
\usepackage{tikz}
\usepackage{pgf-pie} 
\usetikzlibrary{calc}
\usepackage[table,xcdraw]{xcolor}

\newcommand{\commentout}[1]{}

\begin{document}

\date{}

\title{\Large \bf LLM-Driven Kernel Evolution: Automating Driver Updates in Linux}

\commentout{
\author{
{\rm Arina\ Kharlamova}\\
MBZUAI
\and
{\rm Jiawen\ Liu}\\
MBZUAI
\and
{\rm Tianyi\ Zhang}\\
INSPUR
\and
{\rm Xinrui\ Yang}\\
MBZUAI
\and
{\rm Humaid Alqasimi}\\
MBZUAI
\and
{\rm Youcheng Sun}\\
MBZUAI
\and
{\rm Chun Jason Xue}\\
MBZUAI
} 
}

\newcommand{\mbzuaisym}{\textsuperscript{\dag}}

\author{
\begin{minipage}{\textwidth}\centering
{\rm Arina\ Kharlamova}\mbzuaisym \quad
{\rm Jiawen\ Liu}\mbzuaisym \quad
{\rm Tianyi\ Zhang} \quad
{\rm Xinrui\ Yang}\mbzuaisym \quad
{\rm Humaid\ Alqasimi}\mbzuaisym\\[0.9em]
{\rm Youcheng\ Sun}\mbzuaisym \quad
{\rm Chun\ Jason\ Xue}\mbzuaisym\\[0.9em]
\mbzuaisym\;Mohamed bin Zayed University of Artificial Intelligence (MBZUAI)
\end{minipage}
}

\maketitle

\begin{abstract}
Linux kernel evolution breaks drivers through API/ABI changes, semantic shifts, and security-hardening updates. We introduce \textbf{\benchmark}, an executable corpus of kernel$\rightarrow$driver co-evolution cases, and \textbf{\tool}, a closed-loop, LLM-driven system for automating driver maintenance. The system integrates prompt engineering, multi-agent collaboration, static analysis, and iterative validation to ensure that generated patches are not only syntactically correct but also functionally and semantically consistent with kernel conventions. The corpus spans v5.10--v6.10 with \emph{235} validated cases drawn from \emph{612} candidates. In evaluation across \emph{55} cases, \tool achieves \emph{56.4\%} compilation success; QEMU-based boot verification indicates that compiled patches preserve driver initialization in most instances. By releasing \benchmark and tooling, we enable reproducible research and a practical route to continuous, safe co-evolution of drivers with the Linux kernel.
\end{abstract}

\section{Introduction}
\label{sec:intro}

The Linux kernel constitutes the core of the Linux operating system, managing hardware resources and providing fundamental services for all applications  \cite{hayfaasubhi_malallah_b014c2e5, alex_mathai_6fac78f7}. Its widespread deployment across diverse computing environments, from servers to embedded systems, underscores its critical role in modern computing infrastructure  \cite{thong_hoang_955026d7}.

Nevertheless, the continuous updates to the Linux kernel often lead to instability in how different hardware components (drivers) interact with the system. This significantly increases the effort needed for maintenance and can introduce more security vulnerabilities \cite{jukka_ruohonen_4ac57244, guanping_xiao_ce65a421}. These problems happen because changes in the kernel can cause subtle differences in the ways software components communicate (API/ABI divergences), alter the intended meaning of functions (semantic realignments), or introduce new security fixes (hardening patches) that change underlying assumptions about how the system works  \cite{sascha_el_sharkawy_68368dda, shidong_pan_90d3709e, iago_abal_4315d23a}.

\begin{figure}[h!]
  \centering
  \includegraphics[width=\linewidth]{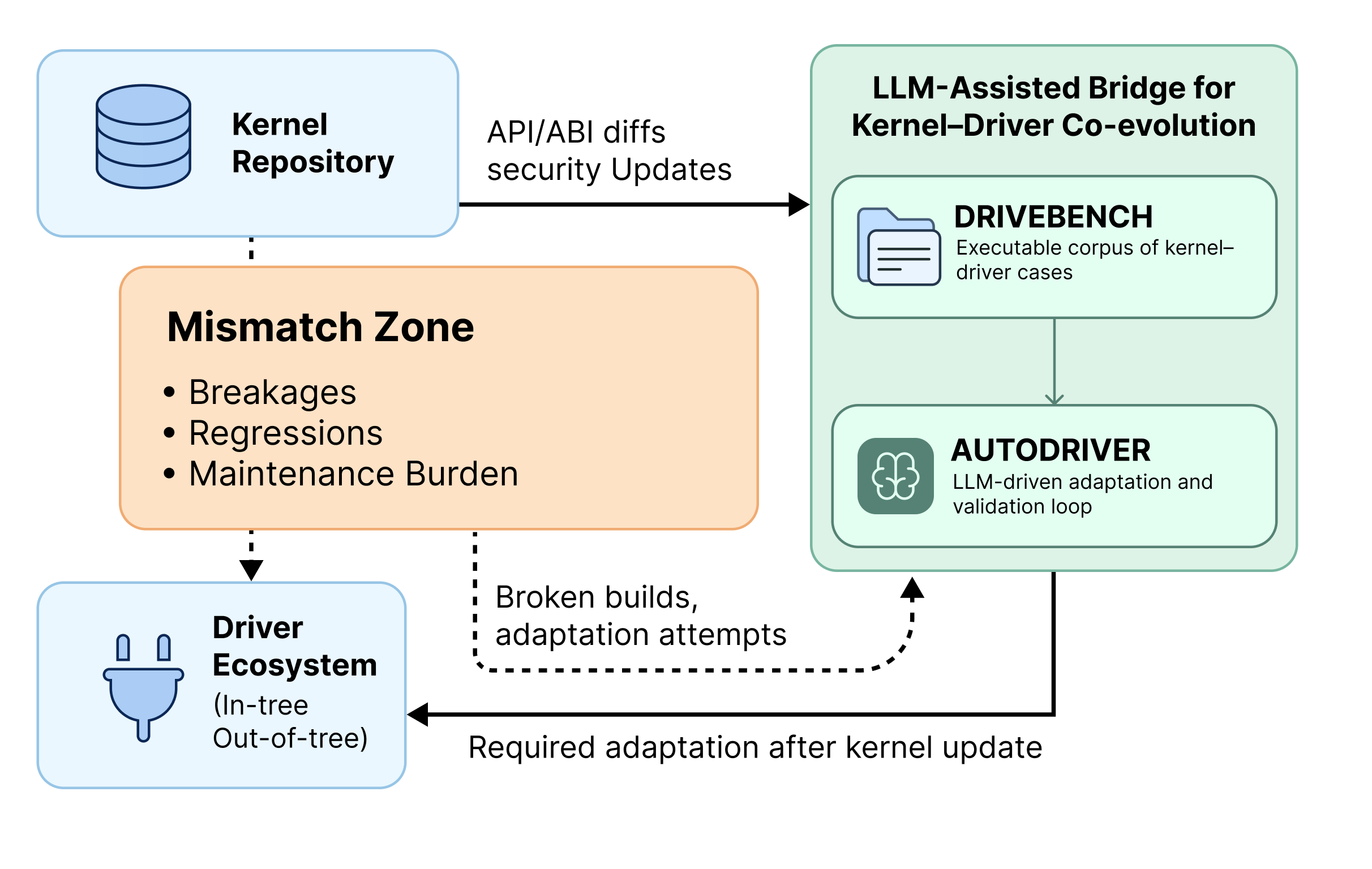}
  \caption[LLM-assisted kernel–driver co-evolution]{
  \textbf{Scheme 1. LLM-assisted co-evolution of Linux kernel and drivers.}
  Kernel evolution introduces \emph{API/ABI, semantic, and security} changes that create a
  \emph{mismatch zone} (e.g., breakages, regressions, and maintenance burden) between the mainline
  kernel and dependent drivers. \textsc{DRIVEBENCH} constructs an \emph{executable corpus} from
  kernel updates and observed breakage episodes; \textsc{AUTODRIVER} consumes this corpus to run an
  \emph{LLM-driven adaptation and validation} loop, returning \emph{validated patches / adapted drivers}
  and thereby closing the gap.
  }
  \label{fig:scheme-coevolution}
\end{figure}

Despite significant advancements in Large Language Models in understanding \cite{j__h_r__jiang_2262f14e, j__h_r__jiang_8bd57e58, nam_huynh_20e27f88, zibin_zheng_a725573f} and generating code \cite{nikhil_pinnaparaju_0d778877, nam_huynh_20e27f88, binyuan_hui_0977b0c6} their full potential remains underutilized in complex domains such as Linux kernel driver evolution. Consequently, kernel driver maintainers continue to grapple with challenges such as incomplete propagation of kernel-side changes  \cite{ridwan_shariffdeen_5bfc4817, x__h__li_34c12965}, insufficient end-to-end validation  \cite{sidney_amani_43106e57}, and latent regressions in performance and security \cite{jukka_ruohonen_4ac57244}. Empirical audits of kernel releases reveal that interface adjustments often lack synchronized driver validation pipelines, leading to runtime inconsistencies and silent semantic drifts \cite{leonid_ryzhyk_7810a3b4}. Sustaining long-term driver viability thus requires not only syntactic adaptation but also preservation of security intent under evolving kernel semantics \cite{yongzhe_huang_23f3dbdd, vitaly_chipounov_ba11f38e}. The underutilization of LLMs in this context underscores an untapped opportunity: \textit{their capabilities in semantic reasoning and large-scale code transformation could potentially mitigate many of these recurring issues if effectively integrated into kernel development workflows.}

However, applying these advancements to highly complex domains like Linux kernel driver evolution presents unique challenges, leading to their underutilization. The Linux kernel is characterized by its multilingual nature, immense size (over 20 million lines of code), critical importance, and highly concurrent operations  \cite{alex_mathai_6fac78f7}. Research indicates that LLMs, while promising, still struggle with aspects such as incomplete code context understanding and accurate migration point identification when dealing with kernel patch migration  \cite{pucheng_dang_54397cc5}. Initial evaluations of LLMs in tasks like Linux kernel crash resolution reveal a significant need for further research to enhance model performance in these intricate software engineering domains  \cite{alex_mathai_6fac78f7}. The intricate and continuously evolving nature of large-scale codebases like the Linux kernel also makes their understanding and analysis challenging, further contributing to the underutilization of LLMs in this area  \cite{haonan_li_94f5114e, yusheng_zheng_6b28a3ff}.

To address the challenges, we propose a unified corpus of executable kernel–driver evolution cases enables reproducible analysis of maintenance dynamics. The \benchmark resource encapsulates commits that modify driver-facing interfaces into executable “case packs,” comprising pre/post commit trees, diffs, build/boot artifacts, and structured metadata. Each case is labeled under a hierarchical taxonomy covering functional, refactoring, and security-driven modifications. The dataset integrates mining and replay utilities that reproduce kernel builds within isolated environments, supporting deterministic validation and differential comparison of driver behavior. The resulting corpus facilitates quantitative study of co-evolution patterns, supporting the derivation of causal relations between interface volatility and adaptation latency.

Automated adaptation is achieved through a \tool that integrates semantic localization, patch synthesis, and executable validation. The localization engine performs dependency-aware impact analysis to constrain the edit region to the minimal affected scope. Patch generation employs taxonomy-conditioned prompting to maintain kernel-idiomatic conventions and type-safety constraints. The refinement pipeline transitions from static verification to dynamic validation through iterative build, emulated boot, and runtime testing cycles. Each stage captures structured logs, configuration states, and test outcomes, enabling post-hoc fault attribution and regression triage. This closed-loop system enforces semantic consistency across both functional and security-preserving updates.

Evaluation across multiple kernel versions demonstrates that taxonomy-aware prompting and staged refinement significantly improve patch correctness and security preservation over baseline generative systems. Comparative analysis across LLM backends isolates the contribution of localization accuracy and taxonomy conditioning to adaptation success rates. These findings provide a quantitative foundation for future kernel-aware model architectures integrating symbolic analysis and empirical validation loops.

\noindent\textbf{Contributions.}
\begin{enumerate}
 \item \textbf{\benchmark: an executable corpus and taxonomy of kernel--driver co-evolution.}
  We release a large-scale, reproducible resource that packages kernel commits affecting drivers into
  \emph{executable case packs} (pre/post trees, diffs, build/boot scripts, logs, tests) and provides a
  hierarchical taxonomy covering functional maintenance (API/ABI migrations, refactors, semantic changes)
  and security-motivated updates (hardening, bug-fix propagation, CVE-linked repairs). We include labeling
  guidelines, and tools for mining, replay, and analysis.
  
  \item \textbf{\tool{} with end-to-end validation for automated driver adaptation.}
  An LLM-driven system that (i) employs a \emph{localization engine} to perform fast, kernel-aware impact analysis and precisely scope edit sites, (ii) synthesizes \emph{minimal} patches, and (iii) iterates a \emph{static-to-dynamic} refinement loop: type/ABI checks and kernel-idiom linters, followed by \emph{build} $\rightarrow$ \emph{QEMU boot} $\rightarrow$ \emph{runtime oracles} that include fast smoke checks, full functional tests, negative (fault-injection) tests, and differential comparisons against a pre-patch or baseline run. The harness records build logs, boot messages, test outcomes, and configuration metadata, and it enforces preservation of security-relevant semantics.
  
  \item \textbf{Unified maintenance results and lessons.}
  We show that one framework handles both functional and security updates. We (i) use taxonomy-aware, category-specific prompts for different classes of kernel--driver updates, (ii) compare multiple LLM backends under identical evaluation, and (iii) develop a fine-tuned model variant trained on case packs and taxonomy signals. We report class- and subtype-stratified outcomes, quantify the effects of localization, staged refinement, and taxonomy-aware prompting via ablations, evaluate cross-release generalization, and analyze failure modes to inform future kernel-aware LLM designs.
\end{enumerate}

\section{Background and Problem Setting}
\label{sec:background}

\definecolor{driveblue}{HTML}{D6EEFF}
\definecolor{driveblue_dark}{HTML}{3E8ED0}

\definecolor{drivegreen}{HTML}{DFFFEF}
\definecolor{drivegreen_dark}{HTML}{3FAE78}

\definecolor{driveorange}{HTML}{FFE7CC}
\definecolor{driveorange_dark}{HTML}{F28C28}

\definecolor{drivered}{HTML}{FFE1E1}
\definecolor{drivered_dark}{HTML}{D35A5A}

\definecolor{drivecyan}{HTML}{D8F7FB}
\definecolor{drivecyan_dark}{HTML}{2BA6C4}

\definecolor{drivepurple}{HTML}{E8E1FF}
\definecolor{drivepurple_dark}{HTML}{7B5CD6}

\definecolor{driveyellow}{HTML}{FFF3C6}
\definecolor{driveyellow_dark}{HTML}{D1A420}

\newcommand\WCthresInner{2}
\newcommand\WCthresOuter{3}
\newcommand\WCtestT[3]{
  \pgfmathparse{\WCpercentage>#3?"#1":"#2"}\pgfmathresult}
\newcommand\WCsmallfont{\tiny} 
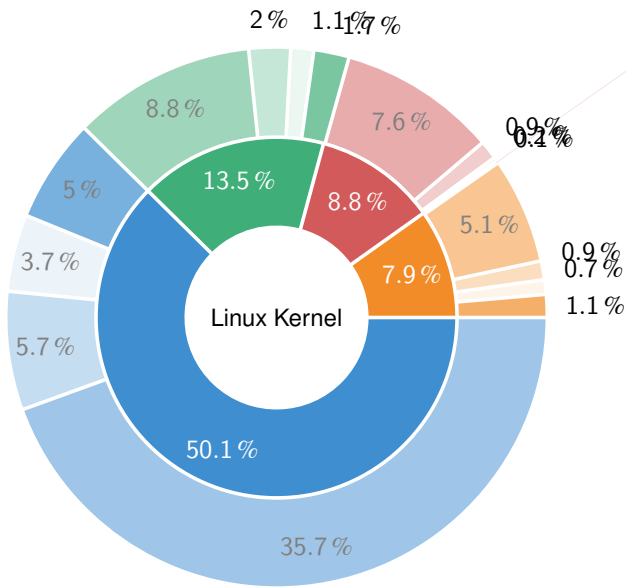
\begin{figure}[b!]
\centering
\begin{tikzpicture}[scale=0.8]
\sffamily

\pgfkeys{
  /wheelchart,
  gap radius=0.03,
  start angle=0,
}

\pgfkeys{
  /wheelchart,
  data=\WCtestT{}{\qty{\WCvarA}{\percent}}{\WCthresInner},
  wheel data=\WCtestT{\qty{\WCvarA}{\percent}}{}{\WCthresInner},
  wheel data style={font=\footnotesize\bfseries},
  wheel data pos=0.58,
  gap=0.03,
}
\wheelchart[
  middle=Linux Kernel,
  radius={1.5}{3.0},
  wheel data style=white
]{%
  50.1/driveblue_dark/drivers,
  13.5/drivegreen_dark/arch,
  8.8/drivered_dark/include,
  7.9/driveorange_dark/fs%
}

\pgfkeys{
  /wheelchart,
  data=\WCtestT{}{\qty{\WCvarA}{\percent}}{\WCthresOuter},
  wheel data=\WCtestT{\qty{\WCvarA}{\percent}}{}{\WCthresOuter},
  wheel data style={font=\WCsmallfont},
  wheel data pos=0.58,                           
}
\wheelchart[
  radius={3.0}{4.5},
  wheel data style=gray
]{%
  35.7/driveblue_dark!50,
  5.7/driveblue_dark!30,
  3.7/driveblue_dark!10,
  5/driveblue_dark!70,
  8.8/drivegreen_dark!50,
  2/drivegreen_dark!30,
  1.1/drivegreen_dark!10,
  1.7/drivegreen_dark!70,
  7.6/drivered_dark!50,
  0.9/drivered_dark!30,
  0.2/drivered_dark!10,
  0.1/drivered_dark!70,
  5.1/driveorange_dark!50,
  0.9/driveorange_dark!30,
  0.7/driveorange_dark!10,
  1.1/driveorange_dark!70%
}
\end{tikzpicture}
\caption{Distribution of code churn across kernel subsystems (Linux v5.10–v6.10). 
Inner ring shows the share of commits per subsystem. 
Outer ring shows the breakdown by number of files modified per commit. 
Colors: blue — \texttt{drivers}, green — \texttt{arch}, red — \texttt{include}, orange — \texttt{fs}.}
\label{fig:kernel_churn_pst}
\end{figure}

The Linux kernel evolves continuously: each release integrates thousands of commits that modify internal APIs, data structures, and synchronization primitives. While such evolution improves performance, scalability, and security, it also imposes a heavy maintenance cost on device drivers, particularly those maintained out-of-tree or by vendors supporting long-lived products. Empirical inspection of versions v5.10–v6.10 shows that driver code accounts for roughly half of all modified lines per release (Fig. \ref{fig:kernel_churn_pst}), a proportion that remains remarkably stable across versions (Table \ref{tab:driver_share}). This concentration of churn indicates that driver co-evolution rather than generic subsystem refactoring constitutes the dominant locus of kernel maintenance effort. Understanding how drivers adapt to shifting kernel semantics is therefore central to sustaining the stability, security, and extensibility of the kernel ecosystem.

\begin{table}[t]
  \centering
  \begin{tabular}{lrr}
    \toprule
    \textbf{Release} & \textbf{Total LoC Modified} & \textbf{Drivers (\%)} \\
    \midrule
    5.11   & 1.06 M     & 72.2 \\
    5.12   & 806.32 K   & 64.9 \\
    5.13   & 902.20 K   & 65.8 \\
    5.14   & 1.13 M     & 75.1 \\
    5.15   & 838.01 K   & 56.7 \\
    5.16   & 931.51 K   & 59.2 \\
    5.17   & 717.30 K   & 54.7 \\
    5.18   & 1.35 M     & 66.9 \\
    5.19   & 1.33 M     & 70.7 \\
    6.0    & 1.65 M     & 66.6 \\
    6.1    & 838.74 K   & 57.0 \\
    6.2    & 1.07 M     & 42.0 \\
    6.3    & 1.13 M     & 55.4 \\
    6.4    & 1.23 M     & 45.1 \\
    6.5    & 716.82 K   & 56.2 \\
    6.6    & 704.81 K   & 53.0 \\
    6.7    & 1.15 M     & 58.7 \\
    6.8    & 844.09 K   & 59.0 \\
    6.9    & 883.84 K   & 58.1 \\
    6.10   & 741.03 K   & 55.6 \\
    \bottomrule
  \end{tabular}
  \caption{Share of modified lines in \texttt{drivers/} across kernel releases.}
  \label{tab:driver_share}
\end{table}

Driver code depends on internal kernel symbols, structure layouts, and initialization sequences that lack formal stability guarantees. Even small changes in the kernel core (for example, field reorderings in \texttt{struct device}, refactored error paths, or updated locking policies) can silently break these dependencies, leading to build-time errors or runtime regressions. Empirical studies have shown that API/ABI drift, missing propagation of interface updates, and misaligned error-handling semantics are among the dominant causes of regressions in the kernel~\cite{iago_abal_4315d23a,ruohonen_fast_2024,leonid_ryzhyk_7810a3b4}.

\subsection{Structural Properties of Kernel Evolution}

\noindent
The Linux kernel does not provide a stable driver interface. In-tree drivers evolve together with the kernel, while out-of-tree drivers must be manually adjusted to track ongoing kernel-side changes. Each release modifies API and ABI boundaries across subsystems such as the driver core, networking, block I/O, and USB. These modifications manifest in three recurring patterns:

\begin{enumerate}
  \item \textit{API/ABI drift}: symbol renaming, function signature changes, or structure reshaping that invalidate existing driver invocations.
  \item \textit{Semantic realignment}: modifications in initialization or teardown order that alter runtime behavior without changing syntax.
  \item \textit{Security hardening divergence}: updates introducing stricter reference counting, pointer checks, or access controls that invalidate assumptions made by legacy code.
\end{enumerate}

Collectively, kernel evolution is non-local, weakly observable, and intent-laden. Automating driver co-evolution thus requires reasoning about distributed semantics under incomplete supervision. Such inconsistencies create a mismatch zone between the evolving kernel (\(K\)) and its dependent driver (\(D\)), resulting in build failures, undefined behavior, or silent runtime faults. Detailed examples of these failures are presented in Section~\ref{sec:motivation}, including the well-documented \texttt{musb\_omap2430} regression case~\cite{lkml_omap2430}.

\subsection{Why Na\"ive LLM Automation Fails}

Large language models (LLMs) exhibit strong surface competence in code generation, yet kernel-scale automation exposes structural mismatches between model abstractions and system reality. 
Unlike user-space software, the Linux kernel operates under weak determinism, non-local semantics, and evolving invariants that are invisible to token-level reasoning. 
The failure of naïve LLM-based automation arises not from the inability to emit compilable code, but from the inability to \emph{control}, \emph{verify}, and \emph{refine} behaviour under such conditions. 
Table~\ref{tab:failures} summarises key failure dimensions and the corresponding abstraction requirements.

\begin{table*}[t]
\centering
\label{tab:failures}
\small
\begin{tabular}{p{0.18\linewidth} p{0.4\linewidth} p{0.35\linewidth}}
\toprule
\textbf{System Property} & \cellcolor{orange!10} \textbf{Observed Automation Failure} & \cellcolor{green!10} \textbf{Required Abstraction} \\
\midrule

\rowcolor{catrow}
\multicolumn{3}{l}{\textit{Contextual and Semantic Failures}} \\

Non-local semantics & Local reasoning fails across interrupt paths, lifecycle callbacks, and deferred contexts; LLMs substitute token-local edits for dependency-aware transformations. & Cross-context semantic graphs, dependency slicing, and evolution-aware prompting. \\[3pt]
Weak oracles & Syntactic success does not imply behavioural correctness; static diffs lack feedback from boot-time or runtime validation. & Multi-stage validation oracles integrating build, boot, and runtime logs within a feedback loop. \\[3pt]
Noisy supervision & Historical commits interleave unrelated intents, producing inconsistent training signals. & Taxonomy-guided supervision and executable case packs exposing pre/post semantics under controlled replay. \\[3pt]

\midrule
\rowcolor{catrow}
\multicolumn{3}{l}{\textit{Specification and Refinement Gaps}} \\

Implicit safety intent & Security-motivated edits (e.g., reference counting, access control) lack explicit specification, leading to semantic erosion. & Intent-preserving objectives capturing invariant semantics beyond syntax (e.g., safety, liveness). \\[3pt]
Iterative refinement & Stateless, single-shot inference ignores compiler and runtime feedback; repeated compilation failures remain unexploited. & Closed-loop dynamic refinement with structured diagnostics and incremental correction. \\
\bottomrule
\end{tabular}
\caption{Structural mismatches between kernel evolution properties and naïve LLM automation, with corresponding required abstractions.}
\end{table*}

\paragraph{Non-local semantics}
Kernel evolution modifies control paths that span multiple layers of the execution hierarchy as interrupt handlers, reference counters, and deferred clean-up routines. 
Since such dependencies are implicit in the call graph and configuration macros, token-based models operate on incomplete local windows. 
Without explicit dependency graphs, LLMs conflate structural edits with behavioural alignment, yielding fragile patches that compile yet fail at runtime.

\paragraph{Weak oracles}
Syntactic correctness is a poor proxy for system validity. 
True correctness emerges only through multi-stage validation: kernel compilation, boot, driver initialisation, and runtime trace analysis. 
Absent such dynamic oracles, LLM automation conflates “buildable” with “correct”, neglecting temporal and semantic invariants that govern driver behaviour.

\paragraph{Noisy supervision}
Historical commits interleave stylistic clean-ups, refactors, and security fixes without consistent metadata. 
When trained on such mixtures, models inherit spurious correlations and cannot distinguish functional adaptation from coincidental edits. 
Structuring supervision through taxonomy-labelled and executable case packs exposes coherent before/after semantics, reducing supervision noise and improving generalisability.

\paragraph{Implicit safety intent} Security-motivated updates encode human intent (e.g., preventing use-after-free, enforcing reference lifetimes) rather than explicit syntactic patterns. Naïve LLMs replicate code shapes while discarding invariant semantics, eroding safety guarantees. Capturing such latent objectives requires reasoning over safety intents (e.g., memory isolation, privilege boundaries, and lifecycle integrity) rather than surface token distributions.

\paragraph{Iterative refinement}
Compilation and QEMU logs provide structured feedback, yet stateless LLMs treat failures as terminal. 
Without feedback incorporation, patch synthesis degenerates into single-shot guessing. 
Embedding compiler diagnostics and runtime traces into the prompting loop converts the process into a feedback-driven optimisation problem, transforming reactive repair into proactive co-evolution.

Collectively, these failure modes expose a structural abstraction gap between \emph{language-level inference} and \emph{system-level determinism}. 
Bridging this gap requires transitioning from static, single-pass generation to a \emph{closed-loop, semantics-aware co-evolution process}, as embodied in our \tool{} framework (Section \ref{sec:autodriver}).

\subsection{Problem Definition and Formalization}
In essence, driver adaptation lags kernel evolution. Every kernel-side modification introduces a latent maintenance debt across its dependent drivers. To close this gap, the adaptation task can be viewed as a synthesis problem. Let \(K\) denote the current kernel source tree and \(K'\) the next version after an update, with the induced change set \(\Delta K = K' - K\). For a driver \(D\) that is compatible with \(K\), the goal is to synthesize a patch \(P\) satisfying:

\[
\begin{aligned}
\texttt{build}(D + P \mid K') &= \textsf{success}, \\
\texttt{run}(D + P \mid K') &\models \Phi_{\text{func}} \wedge \Phi_{\text{sec}}.
\end{aligned}
\]

Here, \(\Phi_{\text{func}}\) and \(\Phi_{\text{sec}}\) represent the functional and security invariants. The functional invariants ensure that the adapted driver reproduces expected behavior across its entry points and callbacks, such as consistent return codes or equivalent traces under regression testing. The security invariants express constraints introduced by kernel hardening updates, including correct reference counting, memory safety, and preservation of isolation guarantees.

\paragraph{Scope and Assumptions}

This study focuses on \textit{source-compatible driver adaptation}, where driver source code is available and can be rebuilt across multiple kernel versions. The framework does not address binary patching, cross-architecture migration, or recovery of missing proprietary dependencies. It assumes reproducible kernel builds and a test harness capable of executing smoke and functional tests for drivers within isolated environments such as QEMU or containerized builds.

\paragraph{Threat Model}

We consider integrity and safety violations that arise from incorrect adaptation rather than from malicious intent. Faulty patches may (i) introduce memory-safety violations, (ii) break locking or reference-counting protocols, or (iii) omit newly required security checks. The framework enforces security-invariant preservation (\(\Phi_{\text{sec}}\)) during validation to prevent these regression-induced vulnerabilities. Broader adversarial threats such as privilege escalation or malicious patch injection are beyond scope.

\subsection{Limitations of Existing Automation}

\noindent
Existing solutions for kernel maintenance, including rule-based porting, symbolic differencing, and static API mapping~\cite{ridwan_shariffdeen_5bfc4817,huang_sok_driver_isolation_2024,fazzini_api_update_2019}, perform poorly in the presence of semantic or cross-subsystem changes. Recent approaches such as \textsc{MigGPT}~\cite{dang_miggpt_2025} demonstrate that large language models can assist with out-of-tree patch migration by encoding contextual structure. However, such systems primarily focus on syntactic patch transfer and lack end-to-end validation or enforcement of security invariants. Consequently, the co-evolution of kernel and drivers remains largely manual, motivating a unified framework that combines executable corpus construction, taxonomy-guided reasoning, and validated synthesis within an LLM-driven workflow.

The challenge addressed in this work is to automatically synthesize an adaptation patch \(P\) that restores compatibility between an updated kernel \(K'\) and its dependent driver \(D\), while maintaining both functional correctness and security integrity across build and runtime validation.

\section{An Motivation Example}
\label{sec:motivation}

Driver breakages following kernel evolution are neither rare nor trivial. A concrete instance occurred during the Linux~4.3 development cycle, when a change in the driver core disrupted initialization semantics in the USB \texttt{musb\_omap2430} controller driver~\cite{lkml_omap2430}. The regression emerged after subtle modifications in the driver-core interface reordered probe-time operations, leaving the OMAP2430 driver unable to register properly on boot. The failure persisted for several days across multiple test builds, prompting a lengthy LKML discussion before maintainers isolated the root cause and proposed a fix. 

This episode exemplifies the fragility of cross-subsystem dependencies: the driver itself had not changed, yet a minor kernel-side refactor silently violated its assumptions. Such regressions typically demand manual diagnosis, patch synthesis, and multi-stage validation across kernel versions---a process that can delay recovery by weeks. Our goal in this work is to formalize and automate this adaptation cycle. By encapsulating cases like the OMAP2430 regression into reproducible, executable “case packs” within \textsc{DriveBench}, we enable deterministic replay, quantitative study of co-evolution latency, and automated patch regeneration through our \textsc{AutoDriver} framework.
\section{\benchmark: Executable Corpus and Taxonomy}
\label{sec:drivebench}

\begin{table*}[t!]
\centering
\label{tab:case-schema}
\small
\renewcommand{\arraystretch}{1.25}
\setlength{\tabcolsep}{6pt}
\begin{tabularx}{\textwidth}{
  >{\ttfamily\raggedright\arraybackslash}p{0.22\textwidth}
  >{\raggedright\arraybackslash}X
}
\toprule
\textbf{Key} &
\cellcolor{orange!15}\textbf{Description} \\
\midrule

\rowcolor{catrow}
\multicolumn{2}{l}{\textit{Core Metadata}} \\

message & Driver commit message. \\
files & List of modified driver source files. \\
patch & Unified diff of the driver update. \\
hash & Driver commit hash. \\[3pt]

\midrule
\rowcolor{catrow}
\multicolumn{2}{l}{\textit{Kernel Association}} \\

kernel-hash & Triggering kernel commit hash. \\

\midrule
\rowcolor{catrow}
\multicolumn{2}{l}{\textit{Classification and Supplementary Data}} \\

type & Category label: \{deprecation/removal, optimisation, regression\}. \\
contents & Full source snapshot of the affected driver files. \\
link (optional) & External reference (e.g., mailing list or vendor patch). \\
description (optional) & Supplementary description or regression report link. \\

\bottomrule
\end{tabularx}
\caption{Unified schema representing a driver update case, linking commit-level artifacts with kernel evolution context and semantic classification.}
\end{table*}

The objective of \benchmark is to operationalise kernel–driver co-evolution as a measurable and reproducible phenomenon.  Rather than a static corpus, \benchmark constitutes a multi-layer data system that integrates three complementary dimensions of change understanding as semantic, analytical, and systemic. \emph{(I) The semantic layer} captures the meaning and intent of code modifications, answering what changes and why it occurs. \emph{(II) The analytical layer} formalises how such changes can be detected, classified, and normalised into structured artefacts suitable for empirical evaluation. \emph{(III) The systemic layer} restores causality and context, linking driver updates to the precise kernel commits that necessitated them. Together, these layers transform raw version-control history into an executable representation of software evolution, supporting both interpretive analysis and deterministic re-execution.

The design adheres to three principles typical of production-grade system datasets: \emph{(I) Traceable provenance}, ensuring that every record can be reconstructed from verifiable Git sources; \emph{(II) Semantic interpretability}, enabling cross-layer reasoning over developer intent and maintenance behaviour; and \emph{(III) Experimental reproducibility}, allowing each change event to be replayed in isolation under controlled build environments.  By aligning these principles, \benchmark acts as the infrastructural substrate for causal and semantic studies of kernel evolution (e.g., analogous in spirit to system benchmarks such as Syzkaller for fuzzing or CSmith for compiler robustness, but oriented toward longitudinal driver co-adaptation).

\begin{figure}[t]
\centering
\includegraphics[width=\linewidth]{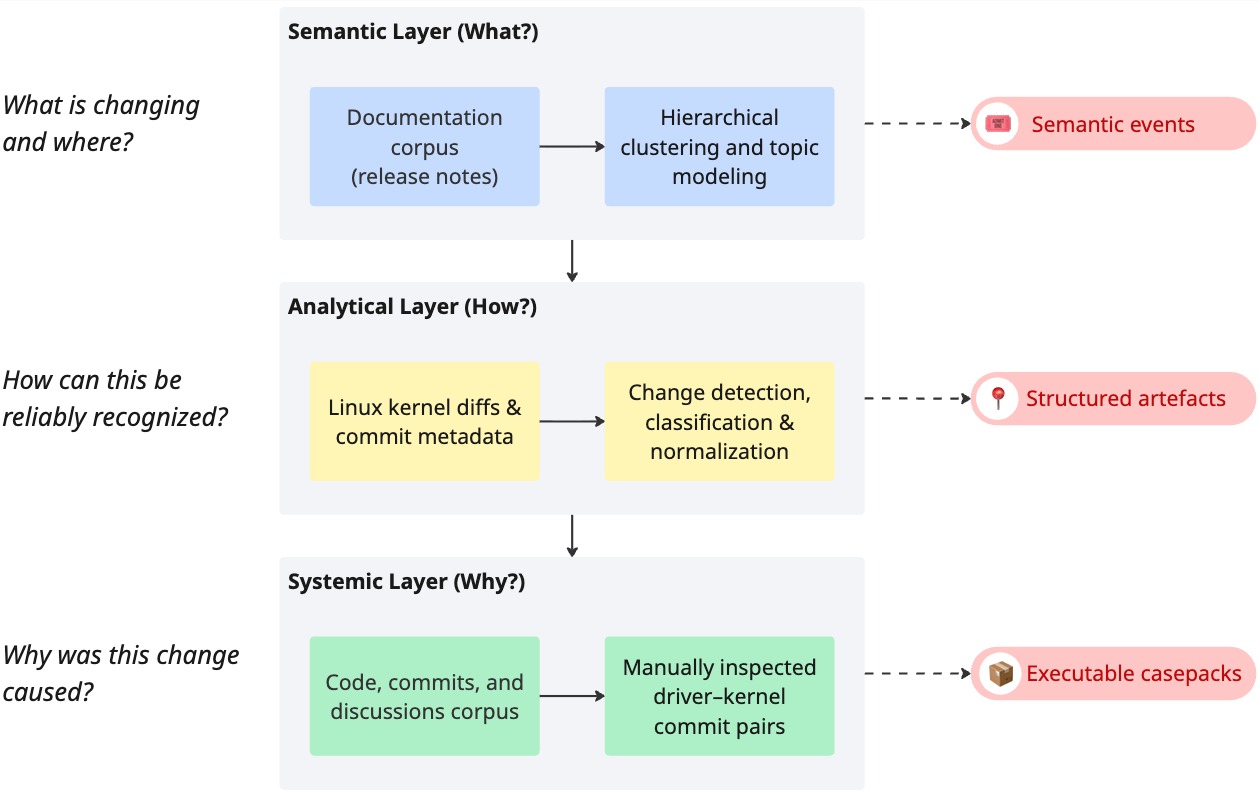}
\caption{Conceptual architecture of \benchmark, aligning the semantic, analytical, and systemic layers into a unified causal data system.}
\label{fig:framework-overview}
\end{figure}

Figure \ref{fig:framework-overview} illustrates the conceptual architecture of \benchmark, showing the interaction between the \emph{semantic} (meaning extraction), \emph{analytical} (structuring and filtering), and \emph{systemic} (causality reconstruction) layers.  Each layer exposes a well-defined interface for data exchange and verification, enabling modular extension and consistent evaluation across future kernel releases.

\subsection{Mining Pipeline}
\label{subsec:mining}

The mining process assembles commits that influence driver-facing interfaces from multiple complementary sources. 
We adopt two acquisition strategies to balance coverage, accuracy, and semantic richness.
(i) The first source is the Linux kernel repository itself. 
Commits are extracted directly from the \texttt{drivers/} subtree across mainline releases, capturing all modifications to in-tree driver code and shared interfaces. 
A filtering stage detects commits that introduce or modify kernel symbols used by drivers, including API migrations, structure changes, and synchronization updates. 
We deduplicate near-identical commits arising from backports or merges and preserve temporal coverage across major kernel releases to enable longitudinal analysis.

(ii) The second source leverages external human-annotated data curated by the \textit{kernelnewbies} project. 
From this documentation corpus we parse structured release summaries to extract commit hashes, natural-language descriptions, categorization tags, and hyperlinks to corresponding mailing-list discussions. 
These records often contain higher-level semantics such as rationale, affected subsystems, and developer intent, providing additional context that is absent from raw diffs.

Subsequently, the two data streams are merged to constitute a unified commit set. Commits derived from the kernel tree guarantee syntactic completeness and the reproducibility of build artifacts, whereas those originating from the \textit{kernelnewbies} project contribute human-readable metadata and a coarse-grained classification. Consequently, the consolidated dataset integrates machine-extracted structural information with community-sourced semantic annotations, thereby enriching both the supervised taxonomy (§\ref{subsec:supervised}) and the semi-supervised clustering pipeline (§\ref{subsec:semisup-delta}).

\subsection{Supervised Taxonomy Construction}
\label{subsec:supervised}

To establish a reliable foundation for commit classification, we initially constructed a supervised taxonomy through manual annotation of kernel commits that influence driver-facing interfaces.

A representative subset of commits underwent selection from the mining pipeline (\S\ref{subsec:mining}) and was subsequently examined by trained annotators. These annotators meticulously inspected each diff, associated metadata, and build context to ascertain the primary maintenance intent. This systematic process yielded a curated anchor set of labeled examples, serving as ground truth for subsequent semi-supervised learning (\S\ref{subsec:semisup-delta}). 

The supervised taxonomy encapsulates diverse evolution intents commonly observed in kernel–driver co-evolution:

\begin{itemize}
    \item \emph{Deprecation:} commits that remove obsolete APIs or mark interfaces as legacy in preparation for migration.
    \item \emph{Transfer:} modifications involving the relocation or refactoring of logic across subsystems or drivers.
    \item \emph{Rename:} symbol and identifier renamings that alter function signatures or structure fields without semantic change.
    \item \emph{Removal:} elimination of redundant functionality or driver backends no longer supported by upstream kernels.
    \item \emph{Simplification:} refactoring to streamline control flow, reduce code duplication, or remove unused abstractions.
    \item \emph{Optimisation:} performance-oriented patches that adjust memory layouts, locking behavior, or I/O scheduling.
    \item \emph{Security:} fixes introducing additional checks, sanitization, or hardening of privilege boundaries.
    \item \emph{Regression:} corrective patches addressing breakages or behavioral deviations introduced by previous kernel updates.
    \item \emph{Hygiene:} stylistic or formatting adjustments that do not alter semantics but improve code readability and maintainability.
\end{itemize}

Each labeled instance is linked to its kernel subsystem, diff metadata, and validation artifacts in DRIVEBENCH. 
Labels were cross-reviewed for consistency, and inter-annotator agreement was periodically audited to ensure quality. 
Single-file changes dominate the supervised dataset, reflecting that most driver evolution tasks involve localized edits, while multi-file cases appear primarily in optimisation and deprecation/removal categories, indicating cross-component refactors or interface-wide clean-ups. 
This supervised anchor dataset defines the semantic space for taxonomy-aware modeling and provides the seed for automated label expansion through the embedding-delta method described in \S\ref{subsec:semisup-delta}.

\subsection{Semi-Supervised Expansion via Code-Embedding Delta}
\label{subsec:semisup-delta}

The supervised taxonomy furnishes high-quality anchors for a restricted subset of commits; however, it is incapable of encompassing the complete heterogeneity of driver-facing modifications at scale. To augment coverage, we implement a semi-supervised clustering methodology that disseminates taxonomy labels via similarity within a joint textual–structural embedding space.

Each commit is represented utilizing two complementary modalities. For the textual component, a BERT-based encoder, fine-tuned on software engineering language, is employed to embed commit messages, associated discussions, and metadata fields from the kernel mailing lists. These embeddings capture human intent, including rationale, maintenance goals, and contextual hints frequently present in commit messages.

For the structural component, we derive a \emph{code-embedding delta}:
\[
v_{\Delta} = \mathrm{normalize}(E_{\text{post}} - E_{\text{pre}}),
\]
where $E_{\text{pre}}$ and $E_{\text{post}}$ are embeddings of the affected code regions before and after the commit. 
This vector captures the direction and magnitude of code evolution, encoding how the change alters structure, dependencies, or semantics.

The message embedding and code-delta representation are fused into a unified vector space, which is trained on manually labeled anchors (\S\ref{subsec:supervised}). Utilizing these anchors as centroids, unlabeled commits are subsequently grouped through similarity-based clustering. Each cluster acquires the label of its closest anchor if its intra-cluster coherence surpasses a predetermined confidence threshold; otherwise, ambiguous clusters persist as unlabeled for subsequent iterations. This iterative process of clustering and propagation establishes a fully automated loop that progressively refines cluster boundaries and extends taxonomy coverage without requiring additional human oversight.

The resultant pseudo-labeled commits, alongside cluster statistics and confidence scores, are incorporated into DRIVEBENCH case packs. These augmented labels furnish enhanced signals for subsequent modules, such as AUTODRIVER’s taxonomy-aware prompting, thereby ameliorating its capacity to discern unforeseen kernel–driver evolution patterns.

\subsection{Case-Pack Format and Tooling}
\label{subsec:casepack}

Each case pack represents a self-contained and executable record of a kernel–driver evolution event. 
The pack structure follows a standardized layout that enables reproducibility and direct integration with AUTODRIVER’s validation pipeline. 
Every pack includes pre- and post-commit source trees, the corresponding diff, and metadata describing the affected files, subsystem, and taxonomy label. 
Build configurations, compiler logs, and runtime traces are embedded alongside minimal replay scripts that reconstruct the build and test process in isolated environments. 
This design allows each pack to be replayed deterministically, enabling quantitative comparison of behavioral differences across kernel versions.

The case-pack format also incorporates structured metadata files that record commit provenance, such as hash, author, and date, along with links to kernel mailing list discussions when available. 
For derived features, each pack stores both textual embeddings from commit messages and code-embedding deltas capturing semantic change magnitude between the pre- and post-commit states. 
These representations are consumed by the semi-supervised clustering pipeline described in §\ref{subsec:semisup-delta} and by taxonomy-aware prompting in AUTODRIVER (§\ref{subsec:supervised}).

Tooling support for DRIVEBENCH automates the creation and replay of these packs. 
The toolchain performs (i) commit extraction and dependency-aware checkout, (ii) deterministic build and boot within containerized or emulated environments, and (iii) structured artifact capture including build results, logs, and runtime test outputs. 
A replayer utility reconstructs any pack end-to-end, while an exporter aggregates metadata and statistics for analysis or model evaluation. 
This modular design ensures that every case—from simple single-file updates to large-scale refactors—can be isolated, rebuilt, and validated under consistent experimental conditions.

\subsection{Dataset Statistics and Release}
\label{subsec:stats}

We report class/subtype and subsystem distributions, human vs.\ auto-labeled ratios, confidence histograms, splits (train/dev/test), and release/licensing details.

\subsection{Dataset Collection and Integration}
\label{sec:dataset}

To evaluate and train LLM-based driver synthesis systems under realistic kernel evolution, we constructed a large-scale dataset capturing driver updates induced by Linux kernel changes. This dataset serves as both a \emph{benchmark} for model evaluation and a \emph{training corpus} for downstream driver code generation.  
We developed a semi-automated pipeline that continuously mines, filters, and structures driver update cases from the Linux kernel repository and regression archives. The resulting dataset spans both \emph{in-tree} and \emph{out-of-tree} driver updates, reflecting the full spectrum of kernel--driver co-evolution behaviors.

\subsubsection{Data Sources and Scope}

Our dataset integrates two complementary data streams:

\begin{itemize}
    \item \emph{In-tree cases:} driver updates directly committed to the Linux mainline repository alongside kernel evolution;
    \item \emph{Out-of-tree cases:} regression fixes and driver updates maintained externally, typically triggered by kernel interface or behavioral changes.
\end{itemize}

This dual-source design captures both proactive adaptation (in-tree) and reactive maintenance (out-of-tree), enabling comprehensive modeling of driver evolution in real-world environments.

\subsubsection{In-tree Mining Pipeline}

We implemented a multi-stage mining pipeline combining LLM-based semantic filtering and automated Git-based extraction.  
The pipeline targets driver commits between \emph{2024-01-01--2025-09-15} from the official Linux repository (\texttt{https://github.com/torvalds/linux}) and proceeds as follows.

\begin{enumerate}
    \item \emph{Commit Retrieval.}  
    A Python crawler enumerates all commits modifying files under \texttt{/drivers/} and extracts metadata including commit hash, commit message, diff and affected paths.

    \item \emph{Initial zero-shot Filtering.}  
    Inspired by the classic four maintenance categories in software engineering, we predefined five corresponding types for Linux kernel commits. Specifically, we further divided the Adaptive category into two subtypes: (1)co-evolution with the kernel, referring to driver changes that adapt to kernel updates; (2)new features and hardware support, referring to commits introducing new functionality or device support. Based on commit messages, we applied the DeBERTa-v3-base-zeroshot-v1 model for zero-shot text classification. The model outputs a \emph{confidence score} in $[0,1]$, representing semantic relevance.

    \item \emph{High-confidence Selection.}  
    Among the commits labeled as \textit{co-evolution with the kernel}, those with confidence scores $\geq 0.5$ were retained, yielding \emph{6,367} candidates for further filtering and analysis.

    \item \emph{Chunked Validation and Classification.}  
    To avoid token overflow, candidates were processed in batches ($\leq$100 commits each). GPT-5 validated adaptation relevance and classified each commit, producing \emph{700} qualified commits.

    \item \emph{Deduplication and Merge Exclusion.}  
    Merge commits were excluded; one invalid hash was discarded, leaving \emph{612} valid driver commits.

    \item \emph{Automated Linkage Construction.}  
    For each valid commit, a Python tool reconstructed the causal linkage between driver and kernel updates.  
    Given a driver commit and its triggering kernel commit hash, the tool retrieved:
    (1) the pre-update driver source,
    (2) the triggering kernel patch, and
    (3) the post-update driver version.  
    All artifacts were stored as structured JSON files grouped by category.

    \item \emph{Manual Verification and Taxonomy Consolidation.}  
    Through iterative GPT-assisted analysis and manual inspection, we initially derived nine fine-grained categories: 
    \textit{deprecation, hygiene, optimisation, removal, rename, security, simplification, transfer,} and \textit{regression}.  
    These categories captured detailed update semantics observed in kernel–driver interactions.

    \item \emph{Final Category Integration.}  
    For our project goal---to evaluate an LLM's ability to \emph{generate and adapt driver code in response to kernel evolution}---we consolidated the above taxonomy into three coarse-grained yet semantically stable categories:
    \begin{itemize}
        \item \emph{API migration:} unified from \textit{deprecation, removal, rename, transfer}, which all correspond to driver changes adapting to API disappearance or signature evolution; and merged from \textit{optimisation, hygiene, simplification, security}, which represent functional improvements or refactoring that preserve behavior while aligning with kernel updates;
        \item \emph{Regression:} retained as-is, denoting driver updates fixing regressions caused by kernel changes.
    \end{itemize}
    This integration simplifies labeling while preserving functional distinction, facilitating both supervised LLM training and quantitative benchmarking.  
    The resulting taxonomy captures the essential driver–kernel adaptation behaviors most relevant to automated driver synthesis and maintenance.
\end{enumerate}

This process produces a structured, semantically grounded dataset directly aligned with the LLM’s target capabilities in our system.

\subsubsection{Out-of-tree Case Acquisition}

To complement in-tree evolution data, we collected out-of-tree regression cases representing driver maintenance external to the mainline kernel.  
We mined two public sources:

\begin{itemize}
    \item \emph{Arch Linux Forum} (\url{https://bbs.archlinux.org/viewforum.php?id=22}): user reports of driver malfunctions following kernel updates;
    \item \emph{Linux Regression Archive} (\url{https://lore.kernel.org/regressions/}): official mailing list documenting kernel regressions, related discussions, and corresponding fixes.
\end{itemize}

Each regression report was traced to (1) the kernel commit introducing the regression and (2) the corresponding driver patch resolving it.  
A Python-based script automatically extracted both commits and their metadata, including diff content, message, and affected files.  
This enabled automatic reconstruction of regression cases analogous to in-tree updates, ensuring schema uniformity across the dataset.

\subsubsection{Unified Case Format}

All driver update cases---in-tree and out-of-tree alike---are normalized into a consistent, machine-readable format.  
Each case is represented as a JSON object following the schema in Table~\ref{tab:case-schema}.

This standardized schema supports structured retrieval and downstream LLM training.  
Each case maintains explicit traceability between driver and kernel commits, ensuring reproducibility and enabling grounded reasoning tasks.

\subsubsection{Tooling and Automation}

To ensure scalability and reproducibility, we built a Python-based mining toolkit that automates data extraction and organization:

\begin{itemize}
    \item \emph{Commit mining and parsing:} automated \texttt{git log} and \texttt{git show} extraction;
    \item \emph{Patch normalization:} unified diff formatting and content hashing;
    \item \emph{Kernel--driver linkage:} automated reconstruction via commit hash correlation;
    \item \emph{Dataset serialization:} structured JSON outputs grouped by category;
    \item \emph{Manual verification support:} visual diff utilities for human-in-the-loop validation.
\end{itemize}

The design minimizes manual intervention while maintaining data fidelity, enabling continuous integration with future kernel releases.

The current dataset captures a part of Linux driver evolution between \emph{2024--2025}, encompassing both proactive in-tree updates and reactive out-of-tree regression fixes.  
In total, it contains:

\begin{itemize}
    \item There are \emph{141} validated in-tree commits and out-of-tree regression cases;
    \item \emph{Three evolution categories:}
        \begin{itemize}
            \item \textit{API migration} --- 141 cases;
            \item \textit{Regression} --- 51 cases.
        \end{itemize}
\end{itemize}

Each case is stored as a structured JSON record linking driver and kernel commits, with complete patch, source, and metadata for reproducibility.  
By combining automated mining, LLM-based semantic classification, and manual curation, the dataset provides a scalable and interpretable foundation for evaluating LLM-driven driver generation systems under real-world kernel evolution dynamics.
\section{\tool: LLM System with End-to-End Validation}
\label{sec:autodriver}

\begin{figure}[t]
    \centering
    \includegraphics[width=\linewidth]{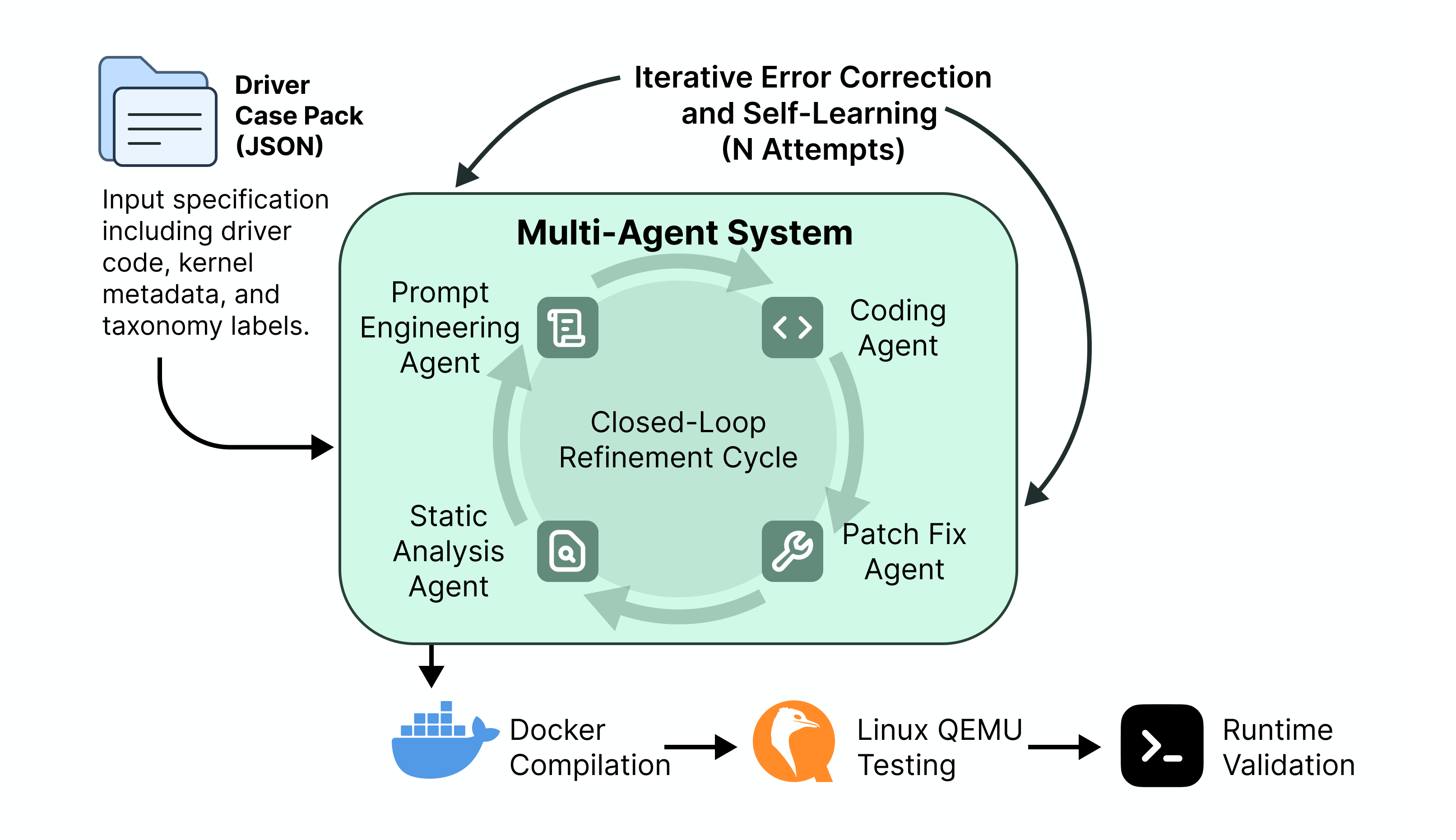}
    \caption{
    \emph{Overview of the Multi-Agent LLM System for Kernel–Driver Adaptation.}
    The framework orchestrates four cooperative agents—\emph{Prompt Engineering}, \emph{Coding}, 
    \emph{Patch Fix}, and \emph{Static Analysis}—within a closed-loop refinement cycle. 
    Starting from a \emph{Driver Case Pack (JSON)} containing driver code, kernel metadata, and taxonomy labels, 
    the agents iteratively synthesize, verify, and repair driver patches under the 
    \emph{Iterative Error Correction and Self-Learning} loop. 
    Upon passing static validation, the pipeline advances through \emph{Docker Compilation}, 
    \emph{Linux QEMU Testing}, and \emph{Runtime Validation}, which collectively ensure 
    functional and security correctness under execution. 
    The resulting patch and runtime report constitute a fully validated output, 
    ready for inclusion in \textsc{DriveBench}.
    }
    \label{fig:multi-agent}
\end{figure}

\tool implements a modular and reproducible workflow for adapting Linux drivers to evolving kernel interfaces. 
The system integrates prompt engineering, multi-agent collaboration, static analysis, and iterative validation to ensure that generated patches are not only syntactically correct but also functionally and semantically consistent with kernel conventions. 
Figure~\ref{fig:multi-agent} presents an overview of the framework, where each component contributes to a closed validation loop between patch synthesis and empirical testing. 

\subsection{Architecture Overview}
AUTODRIVER is designed as a multi-stage pipeline that transforms structured case packs from DRIVEBENCH into validated driver patches. 
The process begins with JSON input containing pre- and post-commit contexts, metadata, and taxonomy labels. 
This input is processed by a coordinated set of agents—Prompt Engineering, Coding, Patch Fix, and Static Analysis—operating within an iterative refinement loop. 
The system automatically performs patch synthesis, format verification, compilation, and runtime testing inside isolated Docker environments. 
Each iteration generates logs and metadata, which are appended to the originating case pack for reproducibility and subsequent learning analysis. 
The multi-agent architecture is inspired by Astra’s cooperative optimization framework but adapted for the constraints of kernel-level software evolution. 

\subsection{Taxonomy-Aware Prompting}
Prompt construction leverages taxonomy information derived from DRIVEBENCH. 
Each update type, such as \textit{deprecation/removal} or \textit{optimisation}, employs a specialized prompt template that conditions the model on relevant patterns and validation requirements. 
Prompts are dynamically assembled from pre- and post-commit code segments, commit messages, and structured metadata. 
This conditioning aligns the generation process with known kernel idioms and style conventions. 
Incorporating taxonomy signals into prompting substantially reduces token redundancy and shortens generation time by focusing the model on relevant code regions. 
Empirical testing shows that generating patches rather than full files cuts token usage by half and reduces API latency from several minutes to under 15 seconds, without compromising code integrity.

\subsection{Localization and Context Extraction}
The localization engine employs \textit{Linux LSP}, an adaptation of the Language Server Protocol specialized for kernel-scale projects. 
Linux LSP performs syntax-based analysis to locate symbol definitions, resolve dependencies, and identify code regions affected by kernel changes. 
It supports cross-file navigation and macro expansion within the Linux build hierarchy, enabling precise mapping between modified interfaces and dependent driver code. 
Instead of supplying complete driver files to the model, the engine extracts local context windows centered around affected functions or structures. 
This targeted extraction reduces input length, minimizes irrelevant edits, and improves synthesis stability. 
Dependency data collected by Linux LSP are passed to downstream stages of AUTODRIVER for guided patch validation and iterative recompilation.

\subsection{Static Analysis and Patch Validation}
Generated patches undergo a two-stage static analysis phase before compilation. 
First, \textit{syntax validation} confirms that each patch adheres to the unified diff format, checks hunk headers and line numbers, and ensures file path consistency. 
Second, \textit{semantic comparison} measures structural similarity to reference patches when available. 
Similarity is computed across multiple levels as file-level mapping, hunk-level matching, and line-level accuracy. 
A configurable threshold (default 0.7) filters low-quality patches from entering the build stage. 
This static validation step improves robustness and reduces wasted compilation attempts, as malformed diffs are rejected early.

\subsection{Compilation and Iterative Fixing}
Once a patch passes static analysis, AUTODRIVER initiates Docker-based compilation within a kernel build environment configured for the corresponding version. 
The system automatically restores the source tree, applies the patch, and launches incremental builds. 
Compilation outcomes drive the iterative repair loop. 
If the build fails, the error analysis module parses compiler diagnostics and returns structured feedback to the Coding Agent, which regenerates or adjusts the patch. 
This error-driven regeneration continues for a configurable number of attempts (default 5–10), allowing the model to refine its output using explicit feedback. 
A two-stage error handling strategy distinguishes between (i) patch application errors, often caused by malformed diffs, and (ii) compilation errors related to API mismatches or dependency resolution. 
This approach has proven effective: up to five iterations capture most recoverable errors, while further attempts yield diminishing.

\subsection{Runtime and Dynamic Validation}
After successful compilation, validated drivers undergo runtime testing in virtualized environments such as QEMU. 
This stage executes smoke and functional tests associated with the driver’s subsystem. 
For hardware-dependent modules, dynamic checks are replaced by static differential analysis that compares pre- and post-execution traces or emulated interactions. 
Future integration with symbolic execution engines such as KLEE and SymDrive is planned to extend coverage to non-virtualizable drivers.

\subsection{Reporting and Continuous Learning}
Each adaptation session produces a detailed report summarizing LLM generation success, compilation success, and runtime validation results across categories. 
Aggregated metrics indicate that patch generation succeeds reliably for all taxonomy types, while compilation rates vary: single-file deprecation and rename updates achieve over 60\% compilation success, whereas multi-file removals remain below 40\%. 
These statistics provide actionable feedback for refining prompts and localization strategies. 
A learning module under development will analyze batch testing results to automatically tune prompting heuristics and improve agent coordination. 
This closed feedback mechanism ensures that AUTODRIVER evolves alongside the kernel itself, maintaining long-term adaptability under continuous kernel change.

\section{Experimental Evaluation}
\label{sec:eval}

We evaluate \tool using the \benchmark to measure the fidelity, structural correctness, and runtime viability of automatically generated Linux kernel patches. Experiments are executed on macOS hosts under Docker with QEMU for runtime checks. Kernel versions span v5.10–v6.10, encompassing common driver update categories such as deprecations, removals, optimisations, and regressions. Two generation engines are tested under identical conditions: \textbf{ChatGPT} and \textbf{DeepSeek}. Both operate within the same iterative regeneration framework that refines patches based on compiler diagnostics until convergence.

\subsection{Infrastructure and Protocol}

Kernel builds require a case-sensitive filesystem, so a dedicated APFS volume was created and mounted into Docker containers. Each build proceeds from a clean kernel tree with recorded logs and artifacts for post-hoc semantic analysis. Every patch pair (reference and generated) is applied to the corresponding baseline file, then evaluated structurally, semantically, and behaviourally. Failed patch applications fall back to hunk reconstruction by extracting context lines, ensuring that no candidate is discarded due to diff misalignment.

\subsection{Tree-sitter–Based AST Scoring}

To evaluate the semantic quality of generated patches, we use \textbf{Tree-sitter}, a production-grade incremental parser, to derive Abstract Syntax Trees (ASTs) for both the \emph{reference} (ground-truth) patch and the \emph{generated} patch. Each patch is applied to the same baseline kernel source file, producing two post-patch versions that can be compared structurally. Tree-sitter parses these files into syntax trees where every node represents a syntactic construct (e.g., \texttt{function\_definition}, \texttt{call\_expression}, \texttt{binary\_expression}) with typed children and a precise text span. The goal is to measure how closely the generated modification mirrors the structural changes introduced by the reference patch.

For each node type \(t\), we count its occurrences in the unmodified baseline (\(N_{\text{base}}(t)\)), in the reference-patched version (\(N_{\text{ref}}(t)\)), and in the generated-patched version (\(N_{\text{gen}}(t)\)). The corresponding change vectors express how each patch alters the baseline:
\[
\Delta_{\text{ref}}(t)=N_{\text{ref}}(t)-N_{\text{base}}(t), \qquad
\Delta_{\text{gen}}(t)=N_{\text{gen}}(t)-N_{\text{base}}(t).
\]
The node-wise similarity between the two patches is then defined as
\[
\mathrm{sim}(t)=1-\frac{|\Delta_{\text{ref}}(t)-\Delta_{\text{gen}}(t)|}{\max(|\Delta_{\text{ref}}(t)|,|\Delta_{\text{gen}}(t)|)}.
\]
Weighting each node type by the magnitude of its structural change yields the aggregated AST similarity:
\[
\mathrm{AST}_{\text{sim}}=\frac{\sum_t w(t)\,\mathrm{sim}(t)}{\sum_t w(t)}, \qquad
w(t)=\max(|\Delta_{\text{ref}}(t)|,|\Delta_{\text{gen}}(t)|).
\]
A value of \(1.0\) indicates that the generated patch makes syntactically identical changes to the reference, while lower values reveal divergence in the kinds or scope of edits. For example, replacing
\texttt{ida\_simple\_get(...,0,0,GFP\_KERNEL)} with
\texttt{ida\_alloc(...,GFP\_KERNEL)} alters the same node categories as the reference patch and therefore yields an AST similarity of \(1.0\); retaining obsolete numeric arguments reduces the score because the added \texttt{number\_literal} nodes differ.

Beyond these quantitative node counts, Tree-sitter queries also extract semantic elements such as modified function definitions, call sites, variable declarations, and macro invocations. Comparing these element sets highlights whether both patches operate on the same logical regions of code. Together, this multi-level AST differencing framework captures the structural and semantic correspondence between generated and reference patches far more accurately than any line-based diff approach.

\subsection{Composite Static Score}

To summarise patch quality across multiple aspects, we compute a weighted static score combining five complementary metrics. Table \ref{tab:metrics} presents the components and weights in a layout suitable for two-column formatting.

\begin{table}[h]
\centering
\caption{Composite static-analysis metrics. Higher is better.}
\label{tab:metrics}
\begin{tabular}{lcc}
\toprule
\textbf{Metric} & \textbf{Symbol} & \textbf{Weight} \\
\midrule
AST Similarity (structural) & $\mathrm{AST}_{\text{sim}}$ & 0.30 \\
Function Accuracy            & $\mathrm{func\_acc}$        & 0.25 \\
Call Accuracy                & $\mathrm{call\_acc}$        & 0.20 \\
Node Accuracy                & $\mathrm{node\_acc}$        & 0.15 \\
Variable Accuracy            & $\mathrm{var\_acc}$         & 0.10 \\
\bottomrule
\end{tabular}
\end{table}

The final score is the weighted sum of these metrics. AST similarity dominates the result since it most strongly reflects semantic alignment.

\subsection{Iterative Regeneration}

An error-driven feedback loop parses compiler diagnostics to guide patch regeneration. Typical recoverable errors, such as missing declarations or incorrect types, are summarised and re-inserted into the next prompt cycle. Most successful patches converge within three to five iterations, beyond which improvements plateau.

\section{Results and Analysis}
\label{sec:results}

The evaluation of \tool demonstrates that kernel patch synthesis can be measured reliably through Tree-sitter–based structural analysis. Both generation backends (e.g., DeepSeek and ChatGPT) were tested under identical iterative regeneration and build conditions, covering 113 Linux driver patches across kernel versions v5.10 to v6.10. The overall goal was to determine whether the generated patches matched the structure, semantics, and build behaviour of their reference updates while maintaining compilation integrity.

DeepSeek compiled successfully in 50 of the 113 test cases, whereas ChatGPT compiled in 25. This difference is consistent across update categories such as optimisation, deprecation, removal, rename, and regression. DeepSeek’s advantage originates from kernel-aware API mappings and template consistency that preserve structural coherence, while ChatGPT exhibits higher stylistic variance. In single-file scenarios, patch application succeeded almost universally, shifting the primary bottleneck from syntax to compilation correctness.

\subsection{Quantitative Results}

Across all metrics, DeepSeek outperforms ChatGPT, achieving an average composite static score of 0.751 compared to 0.697. The distribution of DeepSeek scores is skewed toward higher quality, with a median of 0.783 and a 75th percentile of 0.868, while ChatGPT’s distribution is flatter (median 0.701). Table \ref{tab:metric-breakdown} summarises per-metric averages.

\begin{table}[h]
\centering
\caption{Average metric performance across patch sets.}
\label{tab:metric-breakdown}
\begin{tabular}{lcc}
\toprule
\textbf{Metric} & \textbf{DeepSeek} & \textbf{ChatGPT} \\
\midrule
AST Similarity    & 0.912 & 0.889 \\
Function Accuracy & 0.891 & 0.847 \\
Call Accuracy     & 0.724 & 0.651 \\
Node Accuracy     & 0.438 & 0.401 \\
Variable Accuracy & 0.947 & 0.912 \\
\bottomrule
\end{tabular}
\end{table}

The AST similarity metric, computed from Tree-sitter–derived syntax trees, is the most discriminative measure. It captures structural equivalence rather than textual overlap, providing a sensitive indicator of whether generated patches modify the same syntactic constructs as their references. For example, when both reference and generated patches replace \texttt{ida\_simple\_get} with \texttt{ida\_alloc} and remove identical numeric arguments, the corresponding node-type deltas align perfectly, yielding an AST similarity of 1.0. When the generated patch retains obsolete arguments, node counts diverge, and the similarity drops proportionally. This mechanism quantifies semantic alignment at a level conventional line-based diffs cannot reach.

Statistical testing with Welch’s t-test gives \(t=1.89\), \(p=0.061\), and Cohen’s \(d=0.35\), indicating a small-to-moderate effect close to significance. DeepSeek also produces ten perfect-score patches (17.5 \%) versus three (5.2 \%) for ChatGPT, and nearly twice as many high-quality patches ($\geq$ 0.8). Such differences reflect more stable AST transformations and better adherence to kernel coding idioms.

\subsection{Qualitative Observations}

Inspection of the ten perfect DeepSeek patches shows all involve direct one-to-one API migrations as typical examples include updates of \texttt{devm\_gpio\_chip\_add\_data}, \texttt{devm\_pinctrl\_register}, and the \texttt{ida\_simple\_get} → \texttt{ida\_alloc} substitution. These transformations retain identical function and call patterns, demonstrating that DeepSeek reproduces canonical migration templates with high fidelity. ChatGPT succeeds in simpler renames but often diverges structurally, expanding macros or reordering declarations in ways that reduce AST alignment even when the logic remains correct.

Low-scoring patches share recurring failure modes: incomplete structure updates, partial function migrations, and missing propagation of related changes across dependent code. Representative examples include absent fields in \texttt{ethtool} operation structures and incomplete \texttt{net\_device\_ops} updates. Multi-function or multi-file patches amplify these weaknesses because local regeneration does not yet model inter-file dependencies.

\subsection{Runtime Validation and Regeneration Behaviour}

Runtime smoke tests executed under QEMU confirm that successfully compiled drivers initialise and unload correctly across kernel versions. This indicates that static AST-level similarity correlates strongly with functional stability. Regeneration analysis further shows that most recoverable compiler errors are resolved within the first five iterations, after which success rates plateau. DeepSeek typically converges faster, suggesting that feedback-guided refinement efficiently narrows the edit space, whereas ChatGPT oscillates more between alternative partial fixes.

\subsection{Reproducibility Factors}

Build reproducibility proved sensitive to filesystem configuration. Case-sensitive APFS volumes were mandatory for stable kernel builds on macOS; using standard case-insensitive volumes caused false patch failures due to filename collisions. Docker isolation and deterministic toolchains eliminated host variability, ensuring consistent cross-version results. These controls allow each experiment to be repeated exactly, a crucial property for evaluating stochastic patch generation.

\subsection{Interpretation and Future Outlook}

Taken together, the results demonstrate that Tree-sitter–based AST differencing provides a precise, language-aware measure of patch quality that correlates with human judgement. DeepSeek’s kernel-oriented priors yield structurally coherent patches that closely mirror upstream conventions, while ChatGPT’s general-purpose synthesis remains more exploratory but less consistent. The evaluation also exposes the remaining limitations: AST comparison penalises semantically equivalent yet structurally different rewrites, and static scoring alone cannot confirm runtime equivalence. Incorporating symbolic execution or trace-based validation would address these gaps. Extending the current framework to multi-file updates and subsystem-wide migrations will require coordinated dependency tracking and hierarchical regeneration. Nevertheless, the integration of Tree-sitter analysis, composite scoring, and iterative synthesis establishes a reproducible foundation for automated kernel evolution guided by structural semantics rather than textual similarity.

\section{Discussion}
\label{sec:discussion}

\tool demonstrates that kernel–driver adaptation can be approached as a reproducible, data-driven process rather than a manual engineering task. 
The experimental results confirm that large language models can generate syntactically valid patches consistently across categories, with compilation success reaching over 60\% for deprecation and rename updates. 
The combination of taxonomy-aware prompting, static analysis, and iterative compilation significantly improves robustness compared to single-pass generation. 
Although runtime validation was limited to virtualized testing, the compiled drivers preserved initialization behavior across kernel versions, indicating that the generated patches maintain functional coherence. 
These findings suggest that a hybrid approach combining structured corpus data with adaptive prompting yields tangible benefits for large-scale kernel evolution.

\subsection{Practical Insights from the Evaluation Environment}
The evaluation process uncovered several practical constraints associated with kernel-scale experimentation. 
Running Docker inside macOS required special configuration because the default APFS file system is case-insensitive, while the Linux kernel build process mandates case-sensitive file paths. 
Creating a dedicated APFS volume solved this problem and improved reproducibility across runs. 
This highlights an important lesson for future benchmarks: kernel adaptation pipelines must explicitly account for host system characteristics and container isolation behavior. 
Similarly, using QEMU as a unified runtime environment provided reliable validation for generic drivers, but hardware-dependent subsystems still require hybrid validation combining symbolic and static analysis. 
These experiences indicate that infrastructure fidelity is as crucial as model accuracy for reproducible evaluation.

\subsection{Limitations and Open Challenges}
Despite strong results in single-file commits, the system exhibits reduced performance on multi-file and dependency-heavy updates. 
Such cases often involve interdependent changes across multiple directories or architecture-specific headers, which remain difficult for LLMs to reason about even under iterative refinement. 
The iterative loop mitigates many formatting errors but cannot resolve cases where required kernel symbols were removed or refactored upstream. 
Future work will explore coupling \tool with kernel-aware retrieval systems that dynamically extract API changes from version control or symbol databases to close this context gap. 
Another open issue is test coverage: runtime validation relies primarily on QEMU emulation and cannot capture timing-dependent or hardware-specific regressions. 
Integration with tools such as KLEE and SymDrive will be necessary to expand automated verification for hardware-bound drivers.

\subsection{Toward Continuous Co-evolution}
The modular design of DRIVEBENCH and \tool suggests a pathway toward continuous kernel–driver co-evolution. 
Automated mining, labeling, and replay already enable end-to-end reproducibility; extending this framework into a continuous integration setting would allow real-time driver adaptation as new kernel commits arrive. 
The self-learning mechanism, which aggregates results from batch tests, can serve as a feedback channel to refine prompts and localization heuristics automatically. 
In the long term, coupling this feedback loop with online data from kernel maintainers and mailing lists could transform manual driver maintenance into a semi-autonomous process integrated directly into kernel development workflows.

\subsection{Broader Implications}
The study highlights the feasibility of bringing large language models into safety-critical software maintenance domains traditionally considered beyond their reach. 
Although \tool is designed for Linux, the underlying methodology (e.g., taxonomy-guided reasoning, iterative synthesis, and reproducible replay) applies to other large, evolving codebases such as embedded systems, firmware, and hypervisors. 
By making both the corpus (\benchmark) and the adaptation framework executable and open, this work contributes not only a dataset but a methodology for empirical research on software co-evolution. 
It bridges the gap between experimental LLM reasoning and practical software maintenance, offering a template for reproducible, verifiable AI-assisted engineering in low-level system software.

\section{Related Work}

\subsection{Software Evolution and Variability in the Linux Kernel}

The Linux kernel has long served as a canonical case for examining large-scale software evolution and configurability. Early analyses traced its sustained growth and architectural complexity, revealing continuous expansion shaped by subsystem specialisation and community diversity~\cite{Israeli2010,love2010linux,githubGitHubTorvaldslinux}. Subsequent work framed this process as a manifestation of adaptive, systemic behaviour rather than incidental drift, positioning the kernel as an evolving ecosystem that co-adapts to its development practices.

Building on this foundation, research on kernel variability has elucidated the evolution of its configuration mechanisms and feature model. Lotufo~\emph{et~al.}~\cite{Lotufo2010-oi} reconstructed the historical variability model, showing how configuration options and architectural constraints co-evolve. More recent studies extend this longitudinal perspective: Kuiter~\emph{et~al.}~\cite{Kuiter2025} quantify two decades of feature-model evolution, while Borges~\emph{et~al.}~\cite{Borges2025} provide large-scale datasets enabling reproducible analyses of kernel configurations and performance evolution. Complementary socio-technical investigations interpret Linux development as a form of ``private-collective innovation'', integrating industrial and community contributors within a shared innovation model~\cite{Homscheid2015-mm}.

\subsection{Mining and Understanding Kernel Development Data}

Empirical analyses of Linux kernel development have increasingly relied on large-scale mining of version-control, mailing-list, and issue-tracking data to uncover latent structure in developer activity and software evolution. Early work by Yamauchi~\emph{et~al.}~\cite{Yamauchi2014} demonstrated that clustering commit histories can reveal implementation intents, offering a bridge between low-level change patterns and higher-level maintenance objectives. Subsequent studies have advanced this perspective by applying automated classification and natural language processing techniques to kernel commits. In particular, Dhaouadi~\cite{10.1145/3611643.3617851} introduced a dataset of extracted rationale from commit messages, facilitating fine-grained analyses of developer intent and decision-making processes. Complementary efforts by Gonzalez-Barahona~\emph{et~al.}~\cite{Gonzalez-Barahona2024-ay} extended this line of inquiry through a comprehensive reproduction package for commit classification, establishing methodological baselines for large-scale, reproducible mining of Linux development data.

\subsection{Automation, Visualization, and Educational Tools for Kernel Development}

Recent advances in developer tooling and visual analytics have sought to make Linux kernel development more accessible, reproducible, and cognitively tractable. Visualisation systems such as \emph{ExplorViz}~\cite{Hansen2025} and the interactive framework of Liu~\emph{et~al.}~\cite{Liu2025} employ dynamic graph and dependency visualisations to aid comprehension of subsystem interactions and architectural evolution. These approaches move beyond static documentation, offering real-time perspectives on control flow, module dependencies, and concurrency, thereby facilitating both debugging and design understanding.

Parallel work has focused on automation and pedagogy. Parra~\emph{et~al.}~\cite{Liu2025} present \emph{Kworkflow}, an integrated workflow automation system that streamlines the repetitive tasks of kernel building, patching, and testing, reducing manual configuration overhead. Complementarily, Wen~\emph{et~al.}~\cite{Wen2025} propose \emph{KernelVM}, a browser-based educational environment for teaching kernel development through virtualised experimentation. Together, these efforts represent a shift towards tool-assisted exploration and education, bridging the gap between expert-oriented infrastructure and accessible, interactive environments for both practitioners and learners.

\subsection{Bug Detection, Security, and Intelligent Assistance}

The Linux kernel remains a primary target for automated vulnerability detection and intelligent program analysis due to its size, concurrency, and privileged execution model. Recent advances have combined static reasoning, symbolic execution, and machine learning to improve both vulnerability coverage and diagnostic accuracy. Zhang~\emph{et~al.}~\cite{Zhang2025} introduced a static analysis framework capable of discovering complex cross-entry use-after-free vulnerabilities, revealing latent inter-procedural memory-safety issues beyond the reach of conventional sanitisation tools. Complementarily, Chen~\emph{et~al.}~\cite{Chen2025} proposed \emph{PreXP}, a system that systematically uncovers and exploits security-sensitive kernel objects, broadening the scope of kernel hardening and attack-surface reduction research.

Parallel to static and exploit-oriented analyses, a new line of work explores learning-based and agentic assistance for debugging and maintenance. Zhou~\emph{et~al.}~\cite{Zhou2025} benchmark large language model (LLM) agents on kernel bug localisation tasks, demonstrating their potential yet highlighting substantial reliability gaps. Mathai~\emph{et~al.}~\cite{Mathai2025} present \emph{CrashFixer}, an autonomous crash-resolution agent capable of synthesising candidate patches through structured reasoning over kernel traces. Underpinning such systems, Wei~\emph{et~al.}~\cite{Wei2025} introduce \emph{Sim-CoT}, a supervised implicit chain-of-thought method that improves multi-step reasoning for complex software-analysis pipelines. Collectively, these efforts point toward an emerging synthesis of program analysis, security verification, and intelligent assistance for large-scale systems code.
\section{Conclusion}
\label{sec:conclusion}

This work presents a unified framework for automating driver adaptation in response to kernel evolution. 
Through the combination of \benchmark, a reproducible corpus of executable kernel–driver cases, and \tool, an LLM-driven adaptation system with end-to-end validation, we demonstrate that large language models can effectively support kernel-scale maintenance tasks. 
The integration of taxonomy-guided prompting, dependency-aware localization, static analysis, and iterative refinement enables consistent patch synthesis and compilation success across functional and security-related update categories. 

Our experiments show that \tool achieves complete success in patch generation and over 60\% compilation success on single-file cases, with diminishing returns beyond five iterations. 
The framework’s design ensures that each case can be mined, replayed, and validated deterministically within containerized and emulated environments. 
These results indicate that large language models, when grounded in structured corpora and coupled with programmatic validation, can move beyond purely generative use toward reproducible, verifiable automation for kernel maintenance.

Future extensions of this work include expanding runtime validation through symbolic execution for hardware-bound drivers, integrating continuous feedback loops for model self-tuning, and scaling \benchmark toward a benchmark for evaluating code intelligence on large evolving systems. 
Ultimately, this study outlines a practical path toward continuous kernel–driver co-evolution, bridging the gap between automated reasoning and real-world system reliability in open-source software ecosystems.

\bibliographystyle{plain}
\bibliography{references}

\begin{thebibliography}{10}

\bibitem{githubGitHubTorvaldslinux}
{G}it{H}ub - torvalds/linux: {L}inux kernel source tree --- github.com.
\newblock \url{https://github.com/torvalds/linux}.
\newblock [Accessed 06-10-2025].

\bibitem{iago_abal_4315d23a}
Iago Abal, Claus Brabrand, and Andrzej Wąsowski.
\newblock 42 variability bugs in the linux kernel : A qualitative study.
\newblock {\em Research Portal Denmark}, pages 421--432, 01 2014.

\bibitem{sidney_amani_43106e57}
Sidney Amani, Peter Chubb, Alastair~F. Donaldson, Alexander Legg, Leonid Ryzhyk, and Yanjin Zhu.
\newblock Automatic verification of message-based device drivers.
\newblock {\em arXiv (Cornell University)}, 102:4--17, 11 2012.

\bibitem{Borges2025}
Heraldo Borges, Juliana~Alves Pereira, Djamel~Eddine Khelladi, and Mathieu Acher.
\newblock Linux kernel configurations at scale: A dataset for performance and evolution analysis.
\newblock 5 2025.

\bibitem{Chen2025}
Zuxin Chen, Yaowen Zheng, Hong Li, Siyuan Li, Weijie Wang, Dongliang Fang, Zhiqiang Shi, and Limin Sun.
\newblock Prexp: Uncovering and exploiting security-sensitive objects in the linux kernel.
\newblock {\em IEEE Transactions on Information Forensics and Security}, 2025.

\bibitem{vitaly_chipounov_ba11f38e}
Vitaly Chipounov and George Candea.
\newblock Reverse-engineering drivers for safety and portability.
\newblock pages 1--1, 12 2008.

\bibitem{pucheng_dang_54397cc5}
Pucheng Dang, Di~Huang, Dong Li, Kang Chen, Yuanbo Wen, Qi~Guo, and Xing Hu.
\newblock Miggpt: Harnessing large language models for automated migration of out-of-tree linux kernel patches across versions.
\newblock {\em arXiv (Cornell University)}, 04 2025.

\bibitem{dang_miggpt_2025}
Pucheng Dang, Di~Huang, Dong Li, Kang Chen, Yuanbo Wen, Qi~Guo, Xing Hu, and Ninghui Sun.
\newblock {MigGPT}: Harnessing large language models for automated migration of out-of-tree linux kernel patches across versions.
\newblock arXiv:2504.09474, 2025.

\bibitem{10.1145/3611643.3617851}
Mouna Dhaouadi.
\newblock A data set of extracted rationale from linux kernel commit messages.
\newblock In {\em Proceedings of the 31st ACM Joint European Software Engineering Conference and Symposium on the Foundations of Software Engineering}, ESEC/FSE 2023, page 2187–2188, New York, NY, USA, 2023. Association for Computing Machinery.

\bibitem{sascha_el_sharkawy_68368dda}
Sascha El-Sharkawy, Adam Krafczyk, and Klaus Schmid.
\newblock An empirical study of configuration mismatches in linux.
\newblock 09 2017.

\bibitem{fazzini_api_update_2019}
Mattia Fazzini, Qi~Xin, and Alessandro Orso.
\newblock Automated {API}-usage update for android apps.
\newblock In {\em Proceedings of the 28th {ACM} {SIGSOFT} International Symposium on Software Testing and Analysis (ISSTA '19)}, pages 204--215. ACM, 2019.

\bibitem{Gonzalez-Barahona2024-ay}
Jesus~M Gonzalez-Barahona, Michel Maes, Abhishek Kumar, David Moreno-Lumbreras, and Gregorio Robles.
\newblock Reproduction package for the paper ``classifying linux commits'', 2024.

\bibitem{Hansen2025}
Malte Hansen, Lukas Damerau, Daniel König, and Wilhelm Hasselbring.
\newblock Visualization of the linux kernel with explorviz.
\newblock In {\em 2025 IEEE Working Conference on Software Visualization (VISSOFT)}, pages 117--120. IEEE, 9 2025.

\bibitem{thong_hoang_955026d7}
Thong Hoang, Julia Lawall, Yuan Tian, Richard~J. Oentaryo, and David Lo.
\newblock Patchnet: Hierarchical deep learning-based stable patch identification for the linux kernel.
\newblock {\em IEEE Transactions on Software Engineering}, 47:2471--2486, 11 2019.

\bibitem{Homscheid2015-mm}
Dirk Homscheid, J{\'e}r{\^o}me Kunegis, and Mario Schaarschmidt.
\newblock Private-collective innovation and open source software: Longitudinal insights from linux kernel development.
\newblock In {\em Open and Big Data Management and Innovation}, Lecture notes in computer science, pages 299--313. Springer International Publishing, Cham, 2015.

\bibitem{yongzhe_huang_23f3dbdd}
Yongzhe Huang, Kaiming Huang, Matthew Ennis, Vikram Narayanan, Anton Burtsev, Trent Jaeger, and Gang Tan.
\newblock Sok: Understanding the attack surface in device driver isolation frameworks.
\newblock 01 2024.

\bibitem{huang_sok_driver_isolation_2024}
Yongzhe Huang, Kaiming Huang, Matthew Ennis, Vikram Narayanan, Anton Burtsev, Trent Jaeger, and Gang Tan.
\newblock Sok: Understanding the attack surface in device driver isolation frameworks.
\newblock arXiv:2412.16754, 2024.

\bibitem{binyuan_hui_0977b0c6}
Binyuan Hui, Jian Yang, Zeyu Cui, Jiaxi Yang, Dayiheng Liu, Lei Zhang, Tianyu Liu, Jiajun Zhang, Bowen Yu, Keming Lu, Kai Dang, Yang Fan, Yichang Zhang, An~Yang, Rui Men, Fei Huang, Bo~Zheng, Yibo Miao, Shanghaoran Quan, Yunlong Feng, Xingzhang Ren, Xuancheng Ren, Jingren Zhou, and Junyang Lin.
\newblock Qwen2.5-coder technical report.
\newblock 01 2024.

\bibitem{nam_huynh_20e27f88}
Nam Huynh and Beiyu Lin.
\newblock Large language models for code generation: A comprehensive survey of challenges, techniques, evaluation, and applications.
\newblock 01 2025.

\bibitem{Israeli2010}
Ayelet Israeli and Dror~G. Feitelson.
\newblock The linux kernel as a case study in software evolution.
\newblock {\em Journal of Systems and Software}, 83:485--501, 3 2010.

\bibitem{j__h_r__jiang_2262f14e}
J.-H.R. Jiang, Fan Wang, Jiasi Shen, Seong-Ju Kim, and Sunghun Kim.
\newblock A survey on large language models for code generation.
\newblock {\em arXiv (Cornell University)}, 06 2024.

\bibitem{j__h_r__jiang_8bd57e58}
J.-H.R. Jiang, Fan Wang, Jiasi Shen, Seong-Ju Kim, and Sunghun Kim.
\newblock A survey on large language models for code generation.
\newblock {\em arXiv (Cornell University)}, 06 2024.

\bibitem{Kuiter2025}
Elias Kuiter, Chico Sundermann, Thomas Thüm, Tobias Heß, Sebastian Krieter, and Gunter Saake.
\newblock How configurable is the linux kernel? analyzing two decades of feature-model history – rcr report.
\newblock {\em ACM Transactions on Software Engineering and Methodology}, 8 2025.

\bibitem{haonan_li_94f5114e}
Haonan Li, Hao Yu, Yizhuo Zhai, and Zhiyun Qian.
\newblock Enhancing static analysis for practical bug detection: An llm-integrated approach.
\newblock {\em Proceedings of the ACM on Programming Languages}, 8:474--499, 04 2024.

\bibitem{x__h__li_34c12965}
X.~H. Li, Z.~J. Zhang, Zhiyun Qian, Trent Jaeger, and Chengyu Song.
\newblock An investigation of patch porting practices of the linux kernel ecosystem.
\newblock pages 63--74, 04 2024.

\bibitem{lkml_omap2430}
Linux Kernel~Mailing List.
\newblock usb: musb: omap2430: Fix regression caused by driver core change.
\newblock \url{https://lkml.org/lkml/2015/11/1/202}, 2015.
\newblock Linux Kernel Mailing List discussion, November 2015.

\bibitem{Liu2025}
Hanzhi Liu, Yanyan Jiang, and Chang Xu.
\newblock Understanding the linux kernel, visually.
\newblock In {\em EuroSys 2025 - Proceedings of the 2025 20th European Conference on Computer Systems}, pages 1044--1060. Association for Computing Machinery, Inc, 3 2025.

\bibitem{Lotufo2010-oi}
Rafael Lotufo, Steven She, Thorsten Berger, Krzysztof Czarnecki, and Andrzej Wsowski.
\newblock Evolution of the linux kernel variability model.
\newblock In {\em Software Product Lines: Going Beyond}, Lecture notes in computer science, pages 136--150. Springer Berlin Heidelberg, Berlin, Heidelberg, 2010.

\bibitem{love2010linux}
Robert Love.
\newblock {\em Linux kernel development}.
\newblock Pearson Education, 2010.

\bibitem{hayfaasubhi_malallah_b014c2e5}
HayfaaSubhi Malallah, Subhi R.~M. Zeebaree, Rizgar~R. Zebari, Mohammed A.~M. Sadeeq, Zainab~Salih Ageed, Ibrahim~Mahmood Ibrahim, Hajar~Maseeh Yasin, and Karwan~Jameel Merceedi.
\newblock A comprehensive study of kernel (issues and concepts) in different operating systems.
\newblock {\em Asian Journal of Research in Computer Science}, pages 16--31, 05 2021.

\bibitem{Mathai2025}
Alex Mathai, Chenxi Huang, Suwei Ma, Jihwan Kim, Hailie Mitchell, Aleksandr Nogikh, Petros Maniatis, Franjo Ivančić, Junfeng Yang, and Baishakhi Ray.
\newblock Crashfixer: A crash resolution agent for the linux kernel.
\newblock 5 2025.

\bibitem{alex_mathai_6fac78f7}
Alex Mathai, Chenxi Huang, Petros Maniatis, Aleksandr Nogikh, Franjo Ivancic, Junfeng Yang, and Baishakhi Ray.
\newblock Kgym: A platform and dataset to benchmark large language models on linux kernel crash resolution.
\newblock 01 2024.

\bibitem{shidong_pan_90d3709e}
Shidong Pan, Tianchen Guo, Lihong Zhang, Pei Liu, Zhenchang Xing, and Xiaoyu Sun.
\newblock A large-scale investigation of semantically incompatible apis behind compatibility issues in android apps.
\newblock 01 2024.

\bibitem{nikhil_pinnaparaju_0d778877}
Nikhil Pinnaparaju, Reshinth Adithyan, Duy Phung, Jonathan Tow, James Baicoianu, Ashish Datta, Maksym Zhuravinskyi, Dakota Mahan, Marco Bellagente, Carlos Riquelme, and Nathan Cooper.
\newblock Stable code technical report.
\newblock {\em arXiv (Cornell University)}, 04 2024.

\bibitem{jukka_ruohonen_4ac57244}
Jukka Ruohonen and Adam Alami.
\newblock Fast fixes and faulty drivers: An empirical analysis of regression bug fixing times in the linux kernel.
\newblock 01 2024.

\bibitem{ruohonen_fast_2024}
Jukka Ruohonen and Adam Alami.
\newblock Fast fixes and faulty drivers: An empirical analysis of regression bug fixing times in the linux kernel.
\newblock arXiv:2411.02091, 2024.

\bibitem{leonid_ryzhyk_7810a3b4}
Leonid Ryzhyk, John Keys, Balachandra Mirla, Arun Raghunath, Mona Vij, and Gernot Heiser.
\newblock Improved device driver reliability through hardware verification reuse.
\newblock {\em ACM SIGARCH Computer Architecture News}, 39:133--144, 03 2011.

\bibitem{ridwan_shariffdeen_5bfc4817}
Ridwan Shariffdeen, Xiang Gao, Gregory~J. Duck, Shin~Hwei Tan, Julia Lawall, and Abhik Roychoudhury.
\newblock Automated patch backporting in linux (experience paper).
\newblock pages 633--645, 07 2021.

\bibitem{Wei2025}
Xilin Wei, Xiaoran Liu, Yuhang Zang, Xiaoyi Dong, Yuhang Cao, Jiaqi Wang, Xipeng Qiu, and Dahua Lin.
\newblock Sim-cot: Supervised implicit chain-of-thought.
\newblock 9 2025.

\bibitem{Wen2025}
Elliott Wen, Sean Ma, Paul Denny, Ewan Tempero, Gerald Weber, and Zongcheng Yue.
\newblock Kernelvm: Teaching linux kernel programming through a browser-based virtual machine.
\newblock In {\em SIGCSE TS 2025 - Proceedings of the 56th ACM Technical Symposium on Computer Science Education}, volume~1, pages 1204--1210. Association for Computing Machinery, Inc, 2 2025.

\bibitem{guanping_xiao_ce65a421}
Guanping Xiao, Zheng Zheng, Bo~Jiang, and Yulei Sui.
\newblock An empirical study of regression bug chains in linux.
\newblock {\em IEEE Transactions on Reliability}, 69:558--570, 03 2019.

\bibitem{Yamauchi2014}
Kenji Yamauchi, Jiachen Yang, Keisuke Hotta, Yoshiki Higo, and Shinji Kusumoto.
\newblock Clustering commits for understanding the intents of implementation.
\newblock In {\em Proceedings - 30th International Conference on Software Maintenance and Evolution, ICSME 2014}, pages 406--410. Institute of Electrical and Electronics Engineers Inc., 12 2014.

\bibitem{Zhang2025}
Hang Zhang, Jangha Kim, Chuhong Yuan, Zhiyun Qian, and Taesoo Kim.
\newblock Statically discover complex cross-entry use-after-free vulnerabilities in the linux kernel.
\newblock Internet Society, 3 2025.

\bibitem{yusheng_zheng_6b28a3ff}
Yusheng Zheng, Yiwei Yang, Haoqin Tu, and Yuxi Huang.
\newblock Code-survey: An llm-driven methodology for analyzing large-scale codebases.
\newblock 01 2024.

\bibitem{zibin_zheng_a725573f}
Zibin Zheng, Kaiwen Ning, Yanlin Wang, Jingwen Zhang, Dewu Zheng, Mingxi Ye, and Jiachi Chen.
\newblock A survey of large language models for code: Evolution, benchmarking, and future trends.
\newblock {\em arXiv (Cornell University)}, 01 2023.

\bibitem{Zhou2025}
Zhenhao Zhou, Zhuochen Huang, Yike He, Chong Wang, Jiajun Wang, Yijian Wu, Xin Peng, and Yiling Lou.
\newblock Benchmarking and enhancing llm agents in localizing linux kernel bugs.
\newblock 5 2025.

\end{thebibliography}

\end{document}